\documentstyle[11pt]{article}
\begin{document}

\title{Third Order Effect of Rotation on Stellar Oscillations of a $\beta$-Cephei Star}
\author{K. Karami\thanks{E-mail: KKarami@uok.ac.ir}\\
\small{Department of Physics, University of Kurdistan, Pasdaran
St., Sanandaj, Iran}\\\small{Research Institute for Astronomy $\&$
Astrophysics of Maragha (RIAAM), Maragha, Iran}}

\maketitle

\begin{abstract}
Here the effect of rotation up to third order in the angular
velocity of a star on the p, f and g modes is investigated. To do
this, the third-order perturbation formalism presented by Soufi et
al. (1998) and revised by Karami (2008), was used. I quantify by
numerical calculations the effect of rotation on the oscillation
frequencies of a uniformly rotating $\beta$-Cephei star with
12~$M_\odot$. For an equatorial velocity of 90~$\rm km \, s^{-1}$,
it is found that the second- and third-order corrections for
$(l,m)=(5,-4)$, for instance, are of order of 0.07$\%$ of the
frequency for radial order $n=-3$ and reaches up to 0.6$\%$ for
$n=-20$.
\end{abstract} \noindent{Key words.~~~stars: $\beta$ Cephei variables ---
                stars: oscillation -- stars: rotation}

%---------------------------------------------------------------------
\section{introduction}
Pulsating stars on the upper main sequence, and particularly
$\delta$ Scuti and $\beta$ Cephei stars, are rapid rotators as
well as being multimode pulsators. The ratio
$\epsilon=\Omega/\omega$ of the rotation rate $\Omega$ to the
typical frequency of oscillations $\omega$ seen in these stars is
no longer a small quantity as it is for e.g. the Sun. These stars
have typically equatorial velocities $\sim 100\ {\rm km \,
s^{-1}}$, and oscillation periods from half to a few hours which
implies $\epsilon\sim 0.1$, whereas for the Sun $\epsilon\sim
10^{-4}$. In order to achieve the full potential of
asteroseismology for testing of models for upper main sequence
stars a more careful treatment of the effect of rotation on
oscillation frequencies is required.

Rotation not only modifies the structure of the star but also
changes the frequencies of normal modes. It removes mode
degeneracy creating multiplets of modes. If the rotational angular
velocity, $\Omega$, does not have any latitudinal dependence, and
the rotation is sufficiently slow, the multiplets show a
Zeeman-like equidistant structure. At faster rotation rates
non-negligible quadratic effects in $\Omega$ cause the position of
the centroid frequency of multiplets to shift with respect to that
of a non-rotating model of the same star. See Karami et al.
(2003).

Dziembowski \& Goode (1992) derived a formalism for calculating
the effect of differential rotation on normal modes of rotating
stars up to second order. Soufi et al. (1998) extended the
formalism of Dziembowski \& Goode (1992) up to third order for a
rotation profile that is a function of radius only. Soufi et al.
(1998) found that near-degenerate coupling due to rotation only
occurs between modes with either the same degree $l$ (and
different radial orders) or with modes which differ in degree by
2. In general they showed that the total coupling comes from three
distinct contributions: the Coriolis contribution, the
non-spherically-symmetric distortion, and a coupling term which
involves a combination of these two effects.

Result of calculation of frequency corrections up to third order
were presented for models of $\delta$-Scuti stars by Goupil et al.
(2001), Goupil \& Talon (2002), Pamyatnykh (2003), and Goupil et
al. (2004). Daszy\'{n}ska-Daszkiewicz et al. (2002) studied the
effects of mode coupling due to rotation on photometric parameters
(amplitude and phase) of stellar pulsations. They reconfirmed the
conclusion of Soufi et al. (1998) that the most important effect
of rotation is coupling between close frequency modes of spherical
harmonic degree, $l$, differing by 2 and of the same azimuthal
order, $m$.

Reese et al. (2006) studied the effects of rotation due to both
the Coriolis and centrifugal accelerations on pulsations of
rapidly rotating stars by a non-perturbative method. They showed
that the main differences between complete and perturbative
calculations come essentially from the centrifugal distortion.
Su\'{a}rez et al. (2006) obtained the oscillation frequencies
include corrections for rotation up to second order in the
rotation rate for $\delta$ Scuti star models. Karami (2008)
revised the third-order perturbation formalism presented by Soufi
et al. (1998) because of some misprints and missing terms in some
of their equations. Karami (2008) by the help of the revised
formalism, calculated the effect of rotation up to third order on
the oscillation frequencies of a uniformly rotating zero-age
main-sequence star with 12~$M_\odot$. He concluded that for an
equatorial velocity of 100~$\rm km \, s^{-1}$, the second- and
third-order corrections for $(l,m)=(2,2)$, for instance, are of
order of 0.01$\%$ of the frequency for radial order $n=6$ and
reaches up to 0.5$\%$ for $n=14$.

In this paper, I use the third-order perturbation formalism
according to Soufi et al. (1998) and revised by Karami (2008),
hereafter Paper I. I carry out numerical calculations for the
frequency corrections for a $\beta$ Cephei star with mass $M = 12
\, M_{\odot}$, $M_\odot$ being the solar mass. The numerical
results are presented in section 2. Section 3 is devoted to
concluding remarks.
%---------------------------------------------------------------------

\section{Oscillations of a rapidly rotating $\beta$ Cephei star}\label{II19}
In order to calculate the effect of rotation on normal modes, I
consider a uniformly rotating, 12~$M_\odot$, $\beta$ Cephei model
generated by the evolution code of Christensen-Dalsgaard (1982)
(see also Christensen-Dalsgaard \& Thompson 1999). The parameters
of the model are listed at Table \ref{Model-BetaCephei}. The value
of central hydrogen abundance and existence of small convective
core show that the model should be a quite evolved $\beta$-Cephei
star. The behavior of some of equilibrium quantities of the model
against fractional radius, $x=r/R$ are represented in Fig.
\ref{N2-rho-rho22negative-phi22}; It shows that: 1) Close to the
center up to radius $x=0.1$, the star is in a convective regime
where squared buoyancy frequency $N^2<0$ and, outside of this
radius is in a radiative regime where $N^2>0$. There is a sharp
peak in $N^2$ at $x\simeq0.15$. This happens because on one hand
$N^2\propto g_e$ and on the other hand since the model has a very
small convective core, $R_{\rm{conv}}=0.1~R$, hence the gravity,
$g_e$, increases sharply. 2) Spherically symmetric density $\rho$
decreases smoothly from its maximum value to nearly zero near the
surface at $x\simeq 0.4$; 3) The absolute value of the
non-spherically-symmetric correction to the density $\rho_{22}$,
see Eq. (10) in Paper I, does show a sharp peak at $x\simeq0.15$;
4) The absolute value of the non-spherically-symmetric correction
to the gravitational potential $\phi_{22}$, see Eq. (12) in Paper
I, increases smoothly to its maximum value at the surface.

\subsection{Eigenfunctions} The zero-order eigenfunctions are
computed from the zero order eigenvalue problem with the pulsation
code of Christensen-Dalsgaard (see Christensen-Dalsgaard \&
Berthomieu 1991), modified according to Eqs.~(20) to (24) in Paper
I.

\begin{figure}
 \includegraphics{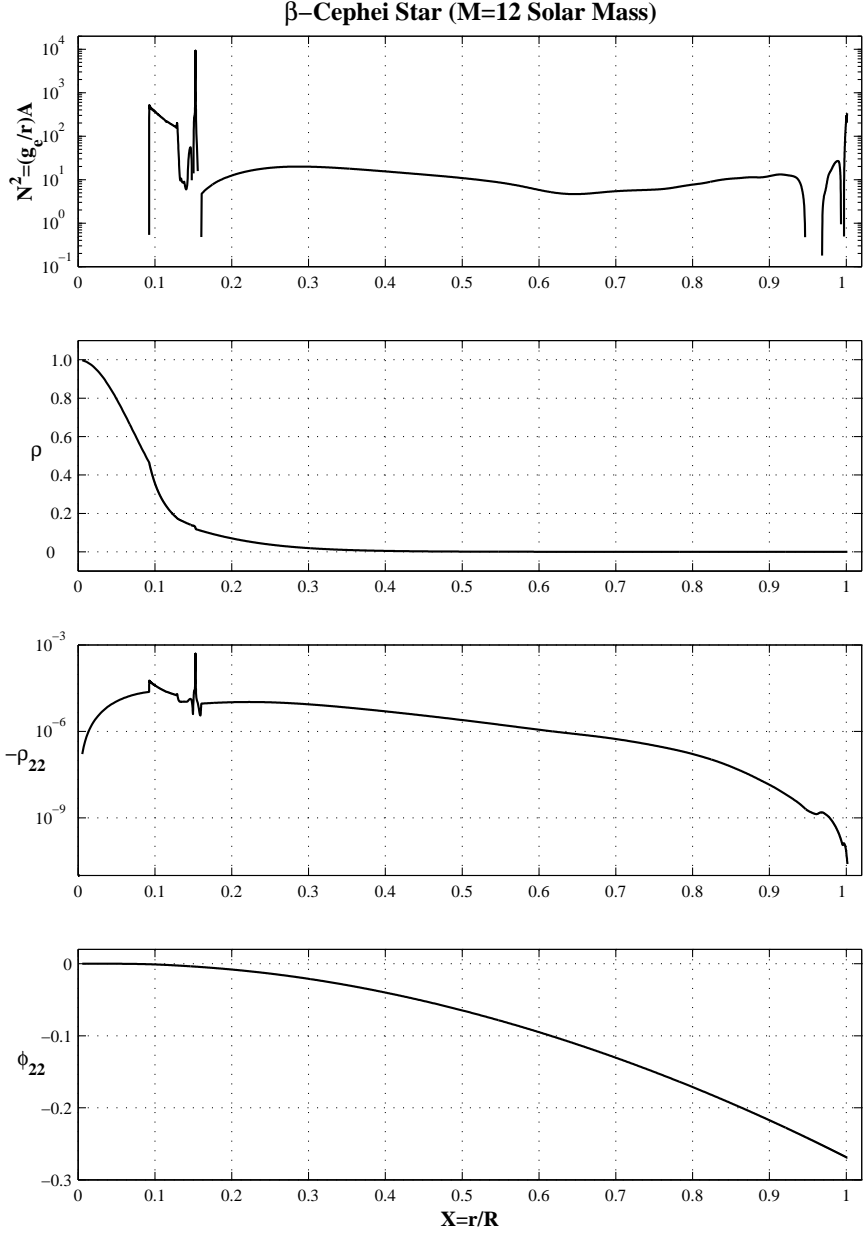}
      \vspace{15.5cm}
      \caption[]{Dimensionless equilibrium quantities including squared
      buoyancy frequency $N^2=(g_{\rm e}/r)A$
      with $A=(1/\Gamma_1)(d\ln p/d \ln x)-(d\ln \rho/d \ln x)$,
      spherically and non-spherically-symmetric contributions
      $\rho$ and $\rho_{22}$ to the density, and
      non-spherically-symmetric gravitational-potential contribution
      $\phi_{22}$, in units of $GM/R^3$, $\rho_{\rm c}$ and
      $R^2\bar{\Omega}^2$, respectively, against fractional radius
      $x=r/R$ for a $\beta$-Cephei star model with $M=12M_{\odot}$.
              }
         \label{N2-rho-rho22negative-phi22}
   \end{figure}

\begin{figure}
 \includegraphics{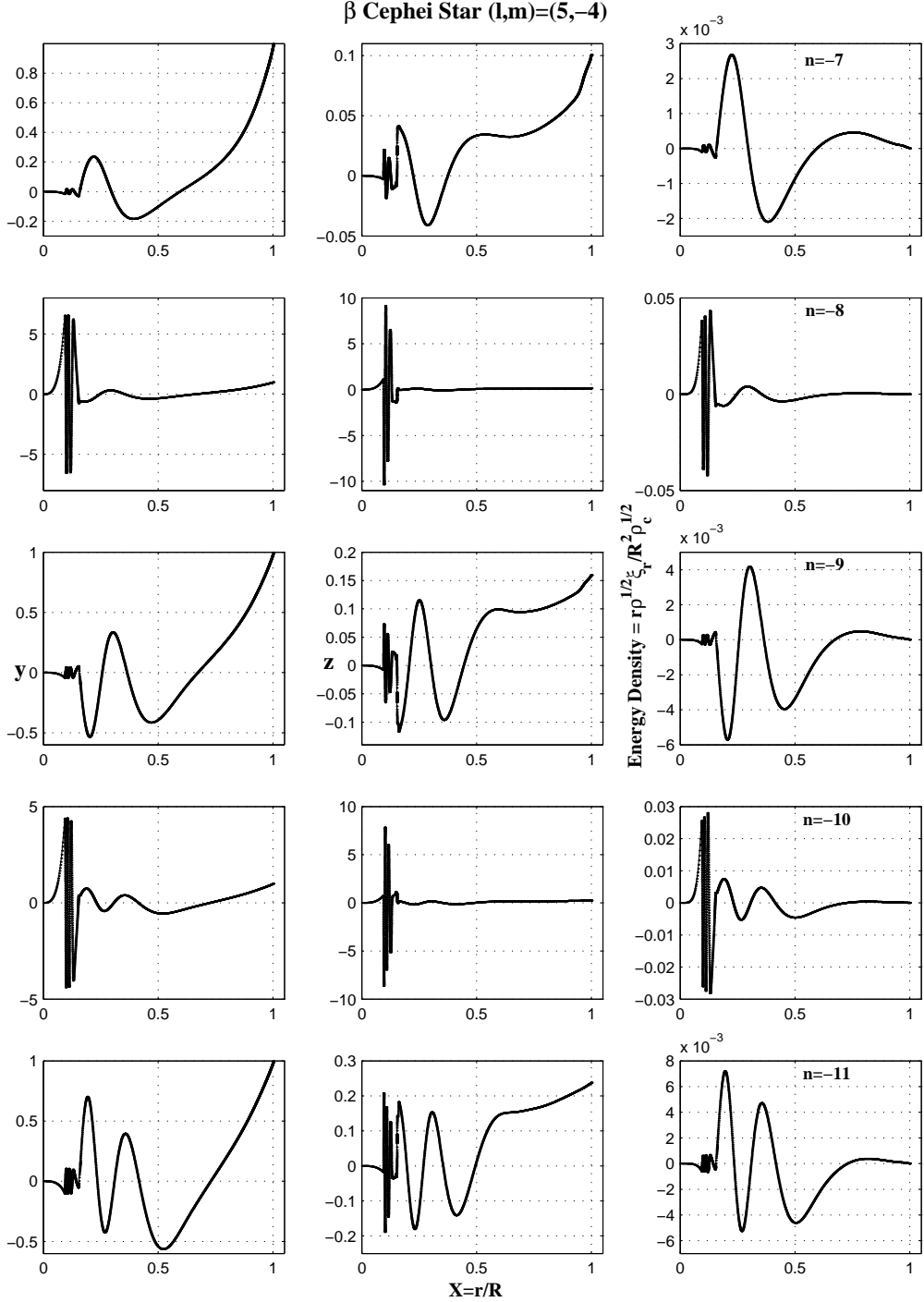}
      \vspace{16.8cm}
\caption[]{Zero-order radial $y$
      (left) and horizontal $z$ components (middle) as well as
      $r\rho^{1/2}\xi_r/(R^2\rho_{\rm c}^{1/2})$, related to the
      energy density (right),
      against fractional radius $x=r/R$ for selected
g-modes with $(l,m)=(5,-4)$
      and $n=(-7,-8,-9,-10,-11)$ for the model
      described in Table \ref{Model-BetaCephei}.
              }
         \label{Fig-Bcephei-yzED-l5m-4-n-7-11}
   \end{figure}

In Fig. \ref{Fig-Bcephei-yzED-l5m-4-n-7-11}, the radial ($y$) and
horizontal ($z$) components of the zero-order poloidal
eigenfunctions as well as $r\rho^{1/2}\xi_r/(R^2\rho_{\rm
c}^{1/2})$ related to the radial energy density, where $\xi_r=ry$
is the radial displacement, are plotted against the fractional
radius $x=r/R$ for the selected modes with $(l,m)$=(5,-4) and
$n$=(-7,...,-11). The modes with $n=(-8,-10)$ are the pure g-modes
and with $n=(-7,-9,-11)$ are the mixed g-modes. For the pure
g-modes, the oscillations are mostly trapped near the center and
also the horizontal amplitude of oscillations are comparable
against the radial component. The horizontal component of buoyancy
force which generates the horizontal amplitude, plays a important
rule during the oscillation of a pure g-mode. At mathematical
point of view a pure g-mode is mostly derived from a vector
potential that its horizontal component has essential contribution
in contrast with the corresponding radial component. This feature
is clear particularly for polytropic models. See Sobouti (1980)
and Sobouti $\&$ Rezania (2001). In the case of mixed g-modes, the
oscillations are trapped between near the center and the middle
part of the star. However the amplitude near the surface is
decayed exponentially. Figure \ref{Fig-Bcephei-yzED-l5m-4-n-7-11}
also shows that: 1) The rapid oscillations in the pure and the
mixed g-modes, occur at $x=0.1$ due to existence of sharp peak in
squared buoyancy frequency $N^2$ (see also Fig.
\ref{N2-rho-rho22negative-phi22}).
 2) Close to the center up to radius
$x=0.1$, where the star is in convective regime ($N^2<0$), the
amplitudes of modes are decayed exponentially
 (see again Fig. \ref{N2-rho-rho22negative-phi22}).

%----------------------------------------------------------------------------------------------
\subsection{Eigenfrequencies and corrections}
The zero-order eigenfrequency, $\sigma_0$, is derived from
numerical solutions of Eqs.~(20) to (24) in Paper I by the
modified pulsation code; note that using the eigensystem in
Eqs.~(20) to (24) the first-order frequency correction,
$\sigma_1$, is implicitly included in $\sigma_0$ (see Eqs.~(15)
and (16) in Paper I). The second- and third-order Coriolis
contributions, ($\sigma_2^{\rm T},~ \sigma_3^{\rm T}$), the
second- and third-order non-spherically-symmetric distortions,
($\sigma_2^{\rm D},~ \sigma_3^{\rm D}$), and the third-order
distortion and Coriolis coupling, $\sigma_3^{\rm C}$, are derived
from numerical integrations of Eqs. (52), (53), (64), (65), and
(69) in Paper I.

In Tables \ref{Table-Bcephei-Sigma0T23D23C-l5m-4} to
\ref{Table-Bcephei-Sigma0T23D23C-l5m4} the results of different
contributions of frequency corrections due to effect of rotation
up to third order are tabulated. In each table the selected p, f
and g modes with $(l,m)=(5,-4)$, $(2,0)$, $(2,1)$, $(2,2)$,
$(2,-1)$, $(5,0)$, $(5,1)$, and $(5,4)$ with $n=(-3, \ldots
,-20)$, $(1, \ldots ,-5)$, $(1, \ldots ,-5)$, $(1, \ldots ,-4)$,
$(1, \ldots ,-5)$, $(-2, \ldots ,-11)$, $(-1, \ldots ,-10)$, and
$(0, \ldots ,-7)$, respectively, are considered. The modes with
$n\geq 1$, $n=0$, and  $n<0$ are labelled by $({\rm p}_1, \ldots
,{\rm p}_n)$, f, and $(g_1,...,g_n)$, respectively.

Tables \ref{Table-Bcephei-Sigma0T23D23C-l5m-4} to
\ref{Table-Bcephei-Sigma0T23D23C-l5m4} show that: 1) The values of
zero order eigenfrequency, $\sigma_0$, and total frequency,
$\sigma_{\rm tot}$, decrease when the radial mode number, $n$,
decreases. 2) The order of magnitudes of ($\sigma_2^{D}$,
$\sigma_3^{D}$) are smaller than ($\sigma_2^{T}$, $\sigma_3^{T}$)
by a factor of $10^{-3}$ to $10^{-1}$. Therefore one can concludes
that the effect of Coriolis forces is dominant with respect to the
centrifugal forces. 3) With increasing $n$, the frequency
correction due to the distortion and the Coriolis coupling
increases and decreases alternatively. 4) For the case of $m=0$,
there is no first and third-order frequency corrections.

In Table \ref{Bcephei-TwoModesCoupling}, the results of third
order frequency corrections for the case of two near degenerate
modes, derived from Eqs. (75)-(76), are tabulated. The coupling
exists only for the two near degenerate modes belonging to the
same $m$ but $l$ differing by $\pm 2$. However there is no any
selection rule for $n$.

Note that in the numerical calculations, there are two substantial
differences between the equations used in Soufi et al. (1998) and
Paper I. In the formulation of Soufi et al. (1998), the density
derivatives are eliminated through an integration by parts and the
resulting surface terms are ignored. The surface terms become
significant for higher-order modes, particularly in the present
model whose atmosphere is relatively thin. The other important
difference which should be noted is that in Soufi et al. (1998)
for computing the third-order correction terms the approximation
$z \simeq y_t/C\sigma_0^2$, which is valid for the non-rotating
case, is used. In Paper I, on the other hand, the exact relation
Eq. (24) is used. The approximation of neglecting the second term
in Eq. (24) everywhere, particularly near the surface, is not
valid. The magnitude of this difference between the two approaches
is more significant than the magnitude of the difference due to of
the surface terms for the present model. If we include the surface
terms in Soufi et al. (1998) and use the approximation $z \simeq
y_t/C\sigma_0^2$ in Paper I, the results of the two numerical
approaches are in good agreement.
%----------------------------------------------------------------------------------------------

\section{Concluding remarks}\label{II20}
The third-order effect of rotation on the p, f and g modes for a
uniformly rotating $\beta$-Cephei star of mass 12~$M_\odot$ has
been investigated. The third-order perturbation formalism
presented by Soufi et al. (1998) and revised by Karami (2008) was
used.
 The zero-order eigenvalue problem was solved by pulsation code modified
in this manner. Numerical calculations of oscillation frequencies
were carried out for our selected model and second- and
third-order frequency corrections due to Coriolis,
non-spherically-symmetric distortion and Coriolis-distortion
coupling were computed. For the case of $m=0$, there is no first
and third-order frequency corrections. Coupling only occurs
between two poloidal modes with
the same $m$ and with $l$ differing by 0 or 2.\\

%-----------------------------------------------------------------------------------------------
\noindent{\bf Acknowledgements} This work was supported by the
Research Institute for Astronomy $\&$ Astrophysics of Maragha
(RIAAM), Maragha, Iran. I wish to thank Prof.
Christensen-Daslgaard for given the model.
%-----------------------------------------------------------------------------------------------------

%----------------------------------------------------------------------------------------------
\clearpage
\begin{table}
\begin{center}
\caption[]{Stellar parameters of a rotating $\beta$-Cephei star in
solar units. $M$, $M_{\rm conv}$, $R$, $R_{\rm conv}$, $p_{\rm
c}$, and $\rho_{\rm c}$ are the total mass, the mass of convective
core, the radius, the radius of convective core, the central
pressure and density, and $\odot$ denotes solar values;
$\sigma_{\bar{\Omega}}$ and $\epsilon$ are the dimensionless mean
angular velocity and the perturbational expansion coefficient;
$T_{\rm dyn}$, $T_{\rm rot}$, and $V_{\rm rot}$ are the dynamical
time scale (free fall time), the equatorial period and velocity,
respectively. For comparison note that $T_{\odot \rm dyn}=0.5~ \rm
h$, $T_{\odot \rm rot}=25~\rm d$, $V_{\odot \rm rot}=2~\rm km \,
s^{-1}$. $X_{\rm{c}}$ is central helium abundance.}
\begin{tabular}{lcc}\hline\noalign{\smallskip}
 $M=12~M_\odot$ & $M_{\rm{conv}}= 0.18~M$ &\\
$R=8.92~R_\odot$ & $R_{\rm{conv}}=0.1~R$  &\\
$P_{\rm{c}}=2.85\times 10^{-1}~P_{\rm{c}\odot}$ & $\rho_{\rm{c}}=5.39\times 10^{-2}~\rho_{\rm{c}\odot}$ &\\
$\sigma_{\bar{\Omega}}=1.78\times10^{-1}$ & $\epsilon=\Omega/\omega=\sigma_{\bar{\Omega}}/2 \pi=2.84\times 10^{-2}$ &\\
$T_{\rm{dyn}}=\sqrt{R^3/GM}=3.41~\rm{h}$ & $T_{\rm{rot}}=2\pi R/V_{\rm{rot}}=4.94~\rm{d}$&\\
$V_{\rm{rot}}=R\Omega=R\sigma_{\bar{\Omega}}/T_{\rm{dyn}}=90~\rm{km/sec}$
& $X_{\rm{c}}= 0.3$&
\\\hline\noalign{\smallskip}
\end{tabular}\\
\label{Model-BetaCephei}
\end{center}
\end{table}
%----------------------------------------------------------------------------------------------
\clearpage
%-------------------------------------------------------------------------------------------------------------
\clearpage
%-----------------------------------------------------------------------------------------------------
\clearpage
\begin{table}
\small{ \caption[]{Values of eigenfrequency $\sigma_0$ (including
the $O(\Omega)$ contribution from rotation), second- and
third-order Coriolis contributions $\sigma_2^{\rm T}$ and
$\sigma_3^{\rm T}$, second- and third-order
non-spherically-symmetric distortions $\sigma_2^{\rm D}$ and
$\sigma_3^{\rm D}$, third-order distortion and Coriolis coupling
$\sigma_3^{\rm C}$, and total frequency $\sigma_{\rm tot}=
\sigma_0+\sigma_{\rm c}$ corrected up to third order, for g modes
in a $\beta$-Cephei star with $M=12M_{\odot}$, $(l,m)=(5,-4)$ and
$n=(-3,...,-20)$. Here $\sigma_{\rm c}=\sigma_2^{\rm
T}+\sigma_3^{\rm T}+\sigma_2^{\rm D} +\sigma_3^{\rm
D}+\sigma_3^{\rm C}$ is the total third-order frequency
correction. All frequencies are in units of
$\sqrt{GM/R^3}=8.16\times 10^{-5}Hz$.}
\begin{center}
\begin{tabular}{rrrcrcrcrcccccc}\hline
\multicolumn{1}{c}{$Mode$}&$n$&\multicolumn{1}{c}{$\sigma_0$}&$\sigma_2^{\rm
T}$&\multicolumn{1}{c}{$\sigma_2^{\rm D}$}&$\sigma_3^{\rm
T}$&\multicolumn{1}{c}{$\sigma_3^{\rm D}$}&$\sigma_3^{\rm
C}$&\multicolumn{1}{c}{$\sigma_{\rm tot}$}&\\\hline
%${\rm g}_2$&-2&         5.3170&   2.5484$\times 10^{-3}$&   -8.1235$\times 10^{-3}$&   3.5065$\times 10^{-4}$&  -1.0402$\times 10^{-3}$&  -1.5568$\times 10^{-4}$&   5.3105&\\
${\rm g}_3$&-3&         4.8665&   2.8674$\times 10^{-3}$&   -3.4879$\times 10^{-5}$&   4.3614$\times 10^{-4}$&  -4.9596$\times 10^{-6}$&   1.3462$\times 10^{-6}$&   4.8698&\\
${\rm g}_4$&-4&        4.4391&   3.4832$\times 10^{-3}$&    -7.1841$\times 10^{-4}$&  6.1277$\times 10^{-4}$&  -1.1948$\times 10^{-4}$&  -1.5434$\times 10^{-5}$&   4.4423&\\
${\rm g}_5$&-5&         4.1259&   2.9132$\times 10^{-3}$&   -2.6820$\times 10^{-3}$&   4.8904$\times 10^{-4}$&  -4.1290$\times 10^{-4}$&  -2.1143$\times 10^{-4}$&   4.1260&\\
${\rm g}_6$&-6&         3.7477&   3.6829$\times 10^{-3}$&   -2.5672$\times 10^{-5}$&   7.2645$\times 10^{-4}$&  -4.7331$\times 10^{-6}$&   1.4759$\times 10^{-6}$&   3.7521&\\
${\rm g}_7$&-7&         3.5272&   4.1565$\times 10^{-3}$&   -2.7378$\times 10^{-4}$&   8.9898$\times 10^{-4}$&  -5.5731$\times 10^{-5}$&   6.8240$\times 10^{-6}$&   3.5319&\\
${\rm g}_8$&-8&        3.0656&   4.4603$\times 10^{-3}$&   -2.1286$\times 10^{-5}$&   1.0752$\times 10^{-3}$&  -4.7965$\times 10^{-6}$&   1.8142$\times 10^{-6}$&   3.0711&\\
${\rm g}_9$&-9&        2.9902&   4.8440$\times 10^{-3}$&   -2.1802$\times 10^{-4}$&   1.2330$\times 10^{-3}$&  -5.2212$\times 10^{-5}$&   2.4225$\times 10^{-5}$&   2.9960&\\
${\rm g}_{10}$&-10&       2.6251&   5.1634$\times 10^{-3}$&   -2.1897$\times 10^{-5}$&   1.4538$\times 10^{-3}$&  -5.7636$\times 10^{-6}$&   2.5355$\times 10^{-6}$&   2.6317&\\
${\rm g}_{11}$&-11&      2.6036&   5.4569$\times 10^{-3}$&   -2.0487$\times 10^{-4}$&   1.5863$\times 10^{-3}$&  -5.5969$\times 10^{-5}$&   3.2563$\times 10^{-5}$&   2.6104&\\
${\rm g}_{12}$&-12&      2.3195&   5.9638$\times 10^{-3}$&   -1.6175$\times 10^{-4}$&   1.9283$\times 10^{-3}$&  -4.9058$\times 10^{-5}$&   3.1539$\times 10^{-5}$&   2.3272&\\
${\rm g}_{13}$&-13&      2.3148&   5.8630$\times 10^{-3}$&   -7.2840$\times 10^{-5}$&   1.8818$\times 10^{-3}$&  -2.1884$\times 10^{-5}$&   1.3520$\times 10^{-5}$&   2.3224&\\
${\rm g}_{14}$&-14&     2.1043&   6.5490$\times 10^{-3}$&   -2.4611$\times 10^{-4}$&   2.3402$\times 10^{-3}$&  -8.2550$\times 10^{-5}$&   5.7688$\times 10^{-5}$&   2.1129&\\
${\rm g}_{15}$&-15&     2.0946&   6.3502$\times 10^{-3}$&   -2.2490$\times 10^{-5}$&   2.2402$\times 10^{-3}$&  -7.4183$\times 10^{-6}$&   4.0019$\times 10^{-6}$&   2.1032&\\
${\rm g}_{16}$&-16&     1.9432&   7.0212$\times 10^{-3}$&   -3.5406$\times 10^{-4}$&   2.7160$\times 10^{-3}$&  -1.2856$\times 10^{-4}$&   9.5675$\times 10^{-5}$&   1.9526&\\
${\rm g}_{17}$&-17&      1.9303&   6.8267$\times 10^{-3}$&   -2.1125$\times 10^{-5}$&   2.6127$\times 10^{-3}$&  -7.5594$\times 10^{-6}$&   4.2244$\times 10^{-6}$&   1.9398&\\
${\rm g}_{18}$&-18&     1.8239&   7.3883$\times 10^{-3}$&   -5.0673$\times 10^{-4}$&   3.0403$\times 10^{-3}$&  -1.9568$\times 10^{-4}$&   1.5051$\times 10^{-4}$&   1.8338&\\
${\rm g}_{19}$&-19&      1.8162&   7.2322$\times 10^{-3}$&   -8.5508$\times 10^{-5}$&   2.9485$\times 10^{-3}$&  -3.2616$\times 10^{-5}$&   2.3645$\times 10^{-5}$&   1.8263&\\
${\rm g}_{20}$&-20&       1.7479&   7.4410$\times 10^{-3}$&   -2.2532$\times 10^{-5}$&   3.1447$\times 10^{-3}$&  -8.9049$\times 10^{-6}$&   5.1061$\times 10^{-6}$&   1.7585&\\
\hline
\end{tabular}\\
\end{center}
\label{Table-Bcephei-Sigma0T23D23C-l5m-4}}
\end{table}
%-----------------------------------------------------------------------------------------------------
%\clearpage
\begin{table}
\small{\caption[]{Same as Table
\ref{Table-Bcephei-Sigma0T23D23C-l5m-4}, for p, f and g modes with
$(l,m)=(2,0)$ and $n=(1, \ldots ,-5)$. There is no $\sigma_3^{\rm
T}$, $\sigma_3^{\rm D}$, and $\sigma_3^{\rm C}=0$ for $m=0$ (see
Eqs.~(53), (65), and (69) in Paper I).}
\begin{center}
\begin{tabular}{rrrcrcrcrccccc} \hline
\multicolumn{1}{c}{$Mode$}&$n$&\multicolumn{1}{c}{$\sigma_0$}&$\sigma_2^{\rm
T}$&\multicolumn{1}{c}{$\sigma_2^{\rm
D}$}&\multicolumn{1}{c}{$\sigma_{\rm tot}$}&\\\hline
$p_{1}$&  1&   5.0537&   5.5170$\times 10^{-3}$&   4.9610$\times 10^{-3}$& 5.0642&\\
$f    $&  0&   4.3575&   6.4970$\times 10^{-3}$&   3.6577$\times 10^{-3}$& 4.3677&\\
$g_{1}$& -1&   4.0537&   6.8344$\times 10^{-3}$&   1.1652$\times 10^{-3}$& 4.0617&\\
$g_{2}$& -2&   3.3908&   8.3406$\times 10^{-3}$&   2.5483$\times 10^{-3}$& 3.4016&\\
$g_{3}$& -3&   2.8396&   9.6665$\times 10^{-3}$&   1.5772$\times 10^{-4}$& 2.8494&\\
$g_{4}$& -4&   2.2565&   1.2362$\times 10^{-2}$&   3.0039$\times 10^{-4}$& 2.2691&\\
$g_{5}$& -5&   1.9548&   1.4006$\times 10^{-2}$&   3.7784$\times 10^{-5}$& 1.9688&\\
\hline
\end{tabular}\\
\end{center}
\label{Table-Bcephei-Sigma0T23D23C-l2m0}}
\end{table}
%-----------------------------------------------------------------------------------------------------
%------------------------------------------------------------------------------------------------------------------------
%\clearpage
\begin{table}
\small{\caption[]{Same as Table
\ref{Table-Bcephei-Sigma0T23D23C-l5m-4}, for p, f and g modes with
$(l,m)=(2,1)$ and $n=(1, \ldots ,-5)$.}
\begin{center}
\begin{tabular}{rrrcrcrcrccccc} \hline
\multicolumn{1}{c}{$Mode$}&$n$&\multicolumn{1}{c}{$\sigma_0$}&$\sigma_2^{\rm
T}$&\multicolumn{1}{c}{$\sigma_2^{\rm D}$}&$\sigma_3^{\rm
T}$&\multicolumn{1}{c}{$\sigma_3^{\rm D}$}&$\sigma_3^{\rm
C}$&\multicolumn{1}{c}{$\sigma_{\rm tot}$}&\\\hline
$p_{1}$&  1&    4.9013&   4.8280$\times 10^{-3}$&   1.3791$\times 10^{-3}$&  -2.4142$\times 10^{-4}$&  -4.2330$\times 10^{-5}$&  -1.0047$\times 10^{-4}$&   4.9072&\\
$ f    $&  0&   4.2006&   5.7720$\times 10^{-3}$&   1.2503$\times 10^{-3}$&  -3.3610$\times 10^{-4}$&  -4.6251$\times 10^{-5}$&  -8.0721$\times 10^{-5}$&   4.2071&\\
$ g_{1}$& -1&   3.9018&   6.1250$\times 10^{-3}$&   3.9277$\times 10^{-4}$&  -3.8355$\times 10^{-4}$&  -1.5264$\times 10^{-5}$&  -2.7455$\times 10^{-5}$&   3.9079&\\
$ g_{2}$& -2&   3.2352&   7.4881$\times 10^{-3}$&   1.1421$\times 10^{-3}$&  -5.6654$\times 10^{-4}$&  -5.3947$\times 10^{-5}$&  -7.4684$\times 10^{-5}$&   3.2431&\\
$ g_{3}$& -3&   2.6895&   9.0188$\times 10^{-3}$&   7.0794$\times 10^{-5}$&  -8.2311$\times 10^{-4}$&  -3.9603$\times 10^{-6}$&  -9.7091$\times 10^{-6}$&   2.6977&\\
$ g_{4}$& -4&   2.1013&   1.2036$\times 10^{-2}$&   1.3298$\times 10^{-4}$&  -1.4124$\times 10^{-3}$&  -9.8793$\times 10^{-6}$&  -3.9305$\times 10^{-5}$&   2.1120&\\
$ g_{5}$& -5&   1.8055&   1.3823$\times 10^{-2}$&   1.8283$\times 10^{-5}$&  -1.8938$\times 10^{-3}$&  -1.5144$\times 10^{-6}$&  -5.6319$\times 10^{-6}$&   1.8174&\\
\hline
\end{tabular}\\
\end{center}
\label{Table-Bcephei-Sigma0T23D23C-l2m1}}
\end{table}
%-----------------------------------------------------------------------------------------------------

%------------------------------------------------------------------------------------------------------------------------
\clearpage
\begin{table}
\small{\caption[]{Same as Table
\ref{Table-Bcephei-Sigma0T23D23C-l5m-4}, for p, f and g modes with
$(l,m)=(2,2)$ and $n=(1, \ldots ,-4)$.}
\begin{center}
\begin{tabular}{rrrcrcrcrccccc} \hline
\multicolumn{1}{c}{$Mode$}&$n$&\multicolumn{1}{c}{$\sigma_0$}&$\sigma_2^{\rm
T}$&\multicolumn{1}{c}{$\sigma_2^{\rm D}$}&$\sigma_3^{\rm
T}$&\multicolumn{1}{c}{$\sigma_3^{\rm D}$}&$\sigma_3^{\rm
C}$&\multicolumn{1}{c}{$\sigma_{\rm tot}$}&\\\hline
$ p_{1}$&  1&    4.7531&   1.5883$\times 10^{-3}$&  -6.2091$\times 10^{-4}$&  -1.1783$\times 10^{-4}$&   3.8294$\times 10^{-5}$&   3.9523$\times 10^{-4}$&   4.7544&\\
$ f    $&  0&    4.0472&   2.1026$\times 10^{-3}$&  -1.2523$\times 10^{-3}$&  -1.8862$\times 10^{-4}$&   9.3932$\times 10^{-5}$&   3.2290$\times 10^{-4}$&   4.0483&\\
$ g_{1}$& -1&    3.7510&   2.2968$\times 10^{-3}$&  -4.5920$\times 10^{-4}$&  -2.2129$\times 10^{-4}$&   3.6960$\times 10^{-5}$&   7.1757$\times 10^{-5}$&   3.7527&\\
$ g_{2}$& -2&    3.0857&   2.6160$\times 10^{-3}$&  -1.9895$\times 10^{-3}$&  -2.9955$\times 10^{-4}$&   1.8946$\times 10^{-4}$&   2.3278$\times 10^{-4}$&   3.0864&\\
$ g_{3}$& -3&    2.5394&   3.4410$\times 10^{-3}$&  -1.2564$\times 10^{-4}$&  -4.8923$\times 10^{-4}$&   1.4920$\times 10^{-5}$&   6.5731$\times 10^{-6}$&   2.5422&\\
$ g_{4}$& -4&    1.9447&   5.0666$\times 10^{-3}$&  -2.3100$\times 10^{-4}$&  -9.7698$\times 10^{-4}$&   3.7477$\times 10^{-5}$&   8.6389$\times 10^{-6}$&   1.9486&\\
\hline
\end{tabular}\\
\end{center}
\label{Table-Bcephei-Sigma0T23D23C-l2m2}}
\end{table}
%-----------------------------------------------------------------------------------------------------

%------------------------------------------------------------------------------------------------------------------------
%\clearpage
\begin{table}
\small{\caption[]{Same as Table
\ref{Table-Bcephei-Sigma0T23D23C-l5m-4}, for p, f and g modes with
$(l,m)=(2,-1)$ and $n=(1, \ldots ,-5)$.}
\begin{center}
\begin{tabular}{rrrcrcrcrccccc} \hline
\multicolumn{1}{c}{$Mode$}&$n$&\multicolumn{1}{c}{$\sigma_0$}&$\sigma_2^{\rm
T}$&\multicolumn{1}{c}{$\sigma_2^{\rm D}$}&$\sigma_3^{\rm
T}$&\multicolumn{1}{c}{$\sigma_3^{\rm D}$}&$\sigma_3^{\rm
C}$&\multicolumn{1}{c}{$\sigma_{\rm tot}$}&\\\hline
$ p_{1}$&  1&   5.2102&   4.4006$\times 10^{-3}$&   3.6153$\times 10^{-3}$&   2.0474$\times 10^{-4}$&   1.1001$\times 10^{-4}$&   3.5096$\times 10^{-5}$&   5.2186&\\
$ f$&  0&       4.5176&   5.1223$\times 10^{-3}$&   2.3647$\times 10^{-3}$&   2.7408$\times 10^{-4}$&   8.4565$\times 10^{-5}$&   2.8493$\times 10^{-5}$&   4.5255&\\
$ g_{1}$& -1&   4.2071&   5.2688$\times 10^{-3}$&   7.9315$\times 10^{-4}$&   3.0026$\times 10^{-4}$&   2.9067$\times 10^{-5}$&   1.7027$\times 10^{-5}$&   4.2135&\\
$ g_{2}$& -2&   3.5517&   6.5049$\times 10^{-3}$&   1.3879$\times 10^{-3}$&   4.4055$\times 10^{-4}$&   6.3833$\times 10^{-5}$&   3.9379$\times 10^{-5}$&   3.5601&\\
$ g_{3}$& -3&   2.9898&   7.1166$\times 10^{-3}$&   8.6542$\times 10^{-5}$&   5.6663$\times 10^{-4}$&   4.3428$\times 10^{-6}$&   1.0038$\times 10^{-5}$&   2.9976&\\
$ g_{4}$& -4&   2.4105&   8.8416$\times 10^{-3}$&   1.6696$\times 10^{-4}$&   8.7025$\times 10^{-4}$&   1.0610$\times 10^{-5}$&   4.4664$\times 10^{-5}$&   2.4204&\\
$ g_{5}$& -5&   2.1045&   9.8081$\times 10^{-3}$&   1.9592$\times 10^{-5}$&   1.1018$\times 10^{-3}$&   1.3914$\times 10^{-6}$&   6.6382$\times 10^{-6}$&   2.1154&\\
\hline
\end{tabular}\\
\end{center}
\label{Table-Bcephei-Sigma0T23D23C-l2m-1}}
\end{table}
%-----------------------------------------------------------------------------------------------------

%------------------------------------------------------------------------------------------------------------------------
%\clearpage
\begin{table}
\small{\caption[]{Same as Table
\ref{Table-Bcephei-Sigma0T23D23C-l2m0}, for g modes with
$(l,m)=(5,0)$ and $n=(-2, \ldots ,-11)$. }
\begin{center}
\begin{tabular}{rrrcrcrcrccccc} \hline
\multicolumn{1}{c}{$Mode$}&$n$&\multicolumn{1}{c}{$\sigma_0$}&$\sigma_2^{\rm
T}$&\multicolumn{1}{c}{$\sigma_2^{\rm
D}$}&\multicolumn{1}{c}{$\sigma_{\rm tot}$}&\\\hline
$ g_{2}$& -2&   4.6430&   6.5692$\times 10^{-3}$&   5.1648$\times 10^{-3}$& 4.6548&\\
$ g_{3}$& -3&   4.1745&   7.4434$\times 10^{-3}$&   4.6698$\times 10^{-5}$& 4.1820&\\
$ g_{4}$& -4&   3.7054&   8.6005$\times 10^{-3}$&   1.0041$\times 10^{-3}$& 3.7150&\\
$ g_{5}$& -5&   3.4917&   8.5212$\times 10^{-3}$&   2.4579$\times 10^{-3}$& 3.5027&\\
$ g_{6}$& -6&   3.0567&   1.0158$\times 10^{-2}$&   3.1727$\times 10^{-5}$& 3.0669&\\
$ g_{7}$& -7&   2.8068&   1.1320$\times 10^{-2}$&   2.2605$\times 10^{-4}$& 2.8184&\\
$ g_{8}$& -8&   2.3747&   1.3073$\times 10^{-2}$&   2.4209$\times 10^{-5}$& 2.3878&\\
$ g_{9}$& -9&   2.2732&   1.3929$\times 10^{-2}$&   1.6712$\times 10^{-4}$& 2.2873&\\
$ g_{10}$&-10&  1.9343&   1.6049$\times 10^{-2}$&   1.9849$\times 10^{-5}$& 1.9504&\\
$ g_{11}$&-11&  1.8916&   1.6674$\times 10^{-2}$&   1.3491$\times 10^{-4}$& 1.9085&\\
\hline
\end{tabular}\\
\end{center}
\label{Table-Bcephei-Sigma0T23D23C-l5m0}}
\end{table}
%-----------------------------------------------------------------------------------------------------

%------------------------------------------------------------------------------------------------------------------------
%\clearpage
\begin{table}
\small{\caption[]{Same as Table
\ref{Table-Bcephei-Sigma0T23D23C-l5m-4}, for g modes with
$(l,m)=(5,1)$ and $n=(-1, \ldots ,-10)$.}
\begin{center}
\begin{tabular}{rrrcrcrcrccccc} \hline
\multicolumn{1}{c}{$Mode$}&$n$&\multicolumn{1}{c}{$\sigma_0$}&$\sigma_2^{\rm
T}$&\multicolumn{1}{c}{$\sigma_2^{\rm D}$}&$\sigma_3^{\rm
T}$&\multicolumn{1}{c}{$\sigma_3^{\rm D}$}&$\sigma_3^{\rm
C}$&\multicolumn{1}{c}{$\sigma_{\rm tot}$}&\\\hline
$ g_{1}$& -1&   5.2735&   5.7438$\times 10^{-3}$&   7.7440$\times 10^{-5}$&  -2.0385$\times 10^{-4}$&  -2.5593$\times 10^{-6}$&  -7.0613$\times 10^{-7}$&   5.2791&\\
$ g_{2}$& -2&   4.4767&   6.5970$\times 10^{-3}$&   2.6340$\times 10^{-3}$&  -2.6521$\times 10^{-4}$&  -9.7601$\times 10^{-5}$&  -1.3530$\times 10^{-4}$&   4.4855&\\
$ g_{3}$& -3&   4.0016&   7.5499$\times 10^{-3}$&   3.9642$\times 10^{-5}$&  -3.5099$\times 10^{-4}$&  -1.7135$\times 10^{-6}$&  -8.7762$\times 10^{-7}$&   4.0088&\\
$ g_{4}$& -4&   3.5245&   8.7660$\times 10^{-3}$&   8.3310$\times 10^{-4}$&  -4.7788$\times 10^{-4}$&  -4.2566$\times 10^{-5}$&  -3.4359$\times 10^{-5}$&   3.5335&\\
$ g_{5}$& -5&   3.3328&   8.6929$\times 10^{-3}$&   1.6871$\times 10^{-3}$&  -4.5513$\times 10^{-4}$&  -8.0691$\times 10^{-5}$&  -1.0396$\times 10^{-4}$&   3.3426&\\
$ g_{6}$& -6&   2.8839&   1.0492$\times 10^{-2}$&   2.6152$\times 10^{-5}$&  -6.7611$\times 10^{-4}$&  -1.5664$\times 10^{-6}$&  -9.6720$\times 10^{-7}$&   2.8938&\\
$ g_{7}$& -7&   2.6261&   1.1821$\times 10^{-2}$&   1.5839$\times 10^{-4}$&  -8.6786$\times 10^{-4}$&  -1.0907$\times 10^{-5}$&  -7.8477$\times 10^{-6}$&   2.6372&\\
$ g_{8}$& -8&   2.2021&   1.3773$\times 10^{-2}$&   1.9341$\times 10^{-5}$&  -1.1620$\times 10^{-3}$&  -1.5166$\times 10^{-6}$&  -9.7065$\times 10^{-7}$&   2.2147&\\
$ g_{9}$& -9&   2.0938&   1.4799$\times 10^{-2}$&   1.0486$\times 10^{-4}$&  -1.3547$\times 10^{-3}$&  -8.9901$\times 10^{-6}$&  -7.3938$\times 10^{-6}$&   2.1073&\\
$ g_{10}$&-10&  1.7617&   1.7257$\times 10^{-2}$&   1.5214$\times 10^{-5}$&  -1.8197$\times 10^{-3}$&  -1.4909$\times 10^{-6}$&  -9.8237$\times 10^{-7}$&   1.7771&\\
\hline
\end{tabular}\\
\end{center}
\label{Table-Bcephei-Sigma0T23D23C-l5m1}}
\end{table}
%-----------------------------------------------------------------------------------------------------

%------------------------------------------------------------------------------------------------------------------------
\clearpage
\begin{table}
\small{\caption[]{Same as Table
\ref{Table-Bcephei-Sigma0T23D23C-l5m-4}, for f and g modes with
$(l,m)=(5,4)$ and $n=(0, \ldots, -7)$.}
\begin{center}
\begin{tabular}{rrrcrcrcrccccc} \hline
\multicolumn{1}{c}{$Mode$}&$n$&\multicolumn{1}{c}{$\sigma_0$}&$\sigma_2^{\rm
T}$&\multicolumn{1}{c}{$\sigma_2^{\rm D}$}&$\sigma_3^{\rm
T}$&\multicolumn{1}{c}{$\sigma_3^{\rm D}$}&$\sigma_3^{\rm
C}$&\multicolumn{1}{c}{$\sigma_{\rm tot}$}&\\\hline
$ f$&  0&   4.9968&   2.6949$\times 10^{-3}$&   2.9857$\times 10^{-3}$&  -3.8428$\times 10^{-4}$&  -3.9355$\times 10^{-4}$&   2.5204$\times 10^{-4}$&   5.0020&\\
$ g_{1}$& -1&   4.7510&   3.0994$\times 10^{-3}$&  -4.5756$\times 10^{-5}$&  -4.8592$\times 10^{-4}$&   6.7139$\times 10^{-6}$&  -7.5470$\times 10^{-7}$&   4.7535&\\
$ g_{2}$& -2&   3.9831&   3.3532$\times 10^{-3}$&   2.5387$\times 10^{-3}$&  -5.9624$\times 10^{-4}$&  -4.1671$\times 10^{-4}$&   3.0860$\times 10^{-4}$&   3.9883&\\
$ g_{3}$& -3&   3.4828&   4.1984$\times 10^{-3}$&  -2.1823$\times 10^{-5}$&  -8.9241$\times 10^{-4}$&   4.3348$\times 10^{-6}$&  -9.4382$\times 10^{-7}$&   3.4861&\\
$ g_{4}$& -4&   2.9964&   4.8000$\times 10^{-3}$&  -1.1922$\times 10^{-4}$&  -1.1736$\times 10^{-3}$&   2.7171$\times 10^{-5}$&   1.2330$\times 10^{-4}$&   3.0000&\\
$ g_{5}$& -5&   2.8464&   4.8577$\times 10^{-3}$&  -1.6612$\times 10^{-4}$&  -1.2257$\times 10^{-3}$&   3.8867$\times 10^{-5}$&   1.0211$\times 10^{-4}$&   2.8500&\\
$ g_{6}$& -6&   2.3658&   6.2203$\times 10^{-3}$&  -1.2587$\times 10^{-5}$&  -1.9446$\times 10^{-3}$&   3.6761$\times 10^{-6}$&  -9.7763$\times 10^{-7}$&   2.3701&\\
$ g_{7}$& -7&   2.0828&   7.6637$\times 10^{-3}$&  -1.8813$\times 10^{-5}$&  -2.8324$\times 10^{-3}$&   6.5530$\times 10^{-6}$&  -9.7497$\times 10^{-6}$&   2.0876&\\
\hline
\end{tabular}\\
\end{center}
\label{Table-Bcephei-Sigma0T23D23C-l5m4}}
\end{table}
%-----------------------------------------------------------------------------------------------------

%------------------------------------------------------------------------------------------------------------------------
%\clearpage
\begin{table}
\small{ \caption[]{Values of zero-order eigenfrequency $\sigma_0$,
total frequency $\sigma_{\pm}$ corrected up to third order, total
third-order frequency correction
$\Delta\sigma_{\pm}=(\sigma_{\pm}-\sigma_0)$ due to rotation and
coupling, expansion coefficients ${\mathcal{A}}_1^{(\pm)}$ and
${\mathcal{A}}_2^{(\pm)}$ normalized to 1, i.e.,
${\mathcal{A}}_1^2+{\mathcal{A}}_2^2=1$, for selected pairs of
near-degenerate poloidal modes in a $\beta$-Cephei star with
$M=12M_{\odot}$. All frequencies are in units of
$\sqrt{GM/R^3}=8.16\times 10^{-5}Hz$.}
\begin{center}
\begin{tabular}{ccccccccccccccc} \hline
$m$&$l$&$n$&coupling&$\sigma_0$&$\sigma_{\pm}$&$\Delta\sigma_{\pm}$&${\mathcal{A}}_1^{(\pm)}$&${\mathcal{A}}_2^{(\pm)}$&\\\hline\hline
0&0&3&$p_3$&5.1186&5.1271&8.4915$\times 10^{-3}$&9.98528$\times 10^{-1}$&5.42353$\times 10^{-2}$&\\
0&2&1&$p_1$&5.0537&5.0640&1.0282$\times 10^{-2}$&5.72211$\times 10^{-2}$&-9.98362$\times 10^{-1}$&\\
\hline
0&2&-2&$g_2$&3.3908&3.4016&1.0888$\times 10^{-2}$&9.99999$\times 10^{-1}$&1.65116$\times 10^{-3}$&\\
0&4&-4&$g_4$&3.4704&3.4795&9.1303$\times 10^{-3}$&8.92439$\times 10^{-3}$&-9.99960$\times 10^{-1}$&\\
\hline
1&3&-3&$g_3$&3.3049&3.3141&9.2414$\times 10^{-3}$&9.98573$\times 10^{-1}$&5.34025$\times 10^{-2}$&\\
1&5&-5&$g_5$&3.3328&3.3426&9.8344$\times 10^{-3}$&4.56655$\times 10^{-2}$&-9.98957$\times 10^{-1}$&\\
%\hline
%2&3&1&$p_1$&5.0380&5.0417&3.7014$\times 10^{-3}$&1.00000&-5.75251$\times 10^{-4}$& \\
%2&5&-1&$g_1$&5.0993&5.1043&5.0349$\times 10^{-3}$&2.69261$\times 10^{-1}$&9.63067$\times 10^{-1}$&\\
\hline
2&3&-3&$g_3$&3.1457&3.1511&5.3915$\times 10^{-3}$&9.98319$\times 10^{-1}$&5.79619$\times 10^{-2}$& \\
2&5&-5&$g_5$&3.1729&3.1809&8.0497$\times 10^{-3}$&5.62952$\times 10^{-2}$&-9.98414$\times 10^{-1}$&\\
\hline
3&3&-3&$g_3$&2.9905&2.9908&3.3252$\times 10^{-4}$&9.98752$\times 10^{-1}$&4.99354$\times 10^{-2}$& \\
3&5&-5&$g_5$&3.0111&3.0171&5.9922$\times 10^{-3}$&5.81273$\times 10^{-2}$&-9.98309$\times 10^{-1}$&\\
\hline
\end{tabular}\\
\end{center}
\label{Bcephei-TwoModesCoupling} }
\end{table}
%-----------------------------------------------------------------------------------------------------

\end{document}